# Few-shot Bioacoustic Event Detection with Machine Learning Methods

**Leah Chowenhill**
lchowenh@andrew.cmu.edu

**Gaurav Satyanath**
gsatyana@andrew.cmu.edu

**Shubhranshu Singh**
shubhran@andrew.cmu.edu

**Madhav Mahendra Wagh**
mwagh@andrew.cmu.edu

## Abstract

Few-shot learning is a type of classification through which predictions are made based on a limited number of samples for each class. This type of classification is sometimes referred to as a meta-learning problem, in which the model learns how to learn to identify rare cases. We seek to extract information from five exemplar vocalisations of mammals or birds and detect and classify these sounds in field recordings [2]. This task was provided in the Detection and Classification of Acoustic Scenes and Events (DCASE) Challenge of 2021. Rather than utilize deep learning, as is most commonly done, we formulated a novel solution using only machine learning methods. Various models were tested, and it was found that logistic regression outperformed both linear regression and template matching. However, all of these methods over-predicted the number of events in the field recordings.

## 1 Introduction

This project focuses on acoustic event identification using the few-shot learning paradigm. Few-shot learning is a type of classification through which predictions are made based on a limited number of samples for each class. It is valuable because it requires less training data and has a low computational cost, while still enabling rare classes to be correctly identified [4]. Further, few shot learning can be used to label events whose category might not be known or have any yet-known label in large acoustic datasets by observing the first few instances of each category [2]. The main objective is to find reliable algorithms that are capable of dealing with data sparsity, class imbalance and noisy/busy environments [2].

Few-shot bioacoustic event detection is most commonly used for biological research. Such classification can be practically applied to survey animal populations, asses the biodiversity of a geographic area, and track range shifts of animal species [4]. For example, this method has been used to track the population of the coquí llanero frog. This frog was discovered in 2005 and inhabits a single wetland in Puerto Rico. The species is highly endangered and has a unique, high-pitched call which lies near the upper range of human hearing. In 2013, biologists at the University of Puerto Rico installed solar-powered iPods in the frog's habitat and began recording the soundscape for 1 minute, every 10 minutes, for 24 hours a day. The research group then used few-shot learning to detect instances of the frog's call in the recordings [10].

## 2 Related Work

For formulating few-shot sound event detection (SED) problem, various traditional supervised methods are used for this setting as well as meta-learning approaches [7] which are conventionally used to solve few-shot classification problem [1] are explored. Few-shot learning has also been studied for use in object detection using sample processing [11].

One of the methods for few-shot recognition is to seek an internal representation on which to perform nearest neighbour classification. This representation should address the variability and noise inherent in few-shot recognition tasks and be generalizable to unseen classes. In [9], the intermediate representation is learned using the notion of micro-sets. A micro-set is a sample of the available data which contains only one instance of each category. The training set is transformed into a number of one-shot optimization tasks corresponding to each micro-set. The model is optimized on the total loss over all micro-sets.

In [8], a prototypical network is used to learn a prototype for each class using few examples. The distance between a query point to each prototype is used to predict the class. The training is performed using episodes of data that resemble the situation during the test time.

## 3 Datasets

For this problem, we use the dataset provided as part of DCASE 2021 challenge [2]. The dataset consists of audio files and annotations (start time, end time and event label) for each audio file. It is necessary to pre-process this dataset to extract the mel spectrogram features. We performed Per channel energy normalisation (PCEN) on mel frequency spectrogram and used it as input feature. The raw audio is scaled to the range $[-231; 231-1]$ before mel transformation and then PCEN is performed using librosa. We used $hop\_length$ of $256$ (samples), $n\_fft$ was set to $1024$ (samples), and $sampling\_rate$ was set to $22050$. The ouptut of preprocessing are equal length patches of spectrogram features and the corresponding label. The training set is balanced using oversampling. The dataset consists of two subsets: Development and Evaluation set. The Development set further consists of training and validation sets. The following subsections provide further information on each of these datasets:

### 3.1 Training Set

The training set is comprised of different families of animal vocalizations such as Hyaenidae, Herpestidae, Corvidae, and Passarellidae [2].The training set consists of $11$ audio recordings with a total duration of $14$ hours and $20$ minutes. The training set consists of $4686$ annotated events categorized into $19$ classes.

Table 1: Training set statistics

| Statistics | Values |
| --- | --- |
| Number of audio recordings | 11 |
| Total duration | 14 hours and 20 minutes |
| Total classes | 19 |
| Total events | 4686 |

### 3.2 Validation Set

None of the classes represented in the training set are present in the validation set (i.e. there is no overlap). The validation set consists of $8$ audio recordings with a total duration of $5$ hours. The validation set consists of $310$ annotated events categorized into $4$ classes. Each recording in the validation set shall be analyzed individually. The validation set also contains single-class annotations of either positive (POS) or unknown (UNK) values.

Table 2: Validation set statistics

| Statistics | Values |
| --- | --- |
| Number of audio recordings | 8 |
| Total duration | 5 hours |
| Total classes | 4 |
| Total events | 310 |

# 4 Results

## 4.1 Evaluation Metric

The validation set provides the first 5 annotations for the class of interest, and the model is tuned to learn a representation from these samples. The tuned model is then evaluated on its prediction for the rest of the audio file. The prediction contains start and end times for the events that possibly contain the class of interest.

In order to evaluate the prediction, first IoU scores are computed with the ground truth, where a single ground truth event may overlap with multiple predicted events [2]. Then, a bipartite graph matching algorithm is used to identify pairs of ground truth and predicted event with maximal matching. During training, the validation accuracy is used as the evaluation metric. During testing, the F-measure is used as an evaluation metric for the matched pairs. The results of these evaluation metrics for various models are shown in Table 3 and Table 4.

Table 3: Validation Accuracy

| Model | Accuracy |
|---|---|
| DCASE CNN | 98.55 % |
| Logistic Baseline | 90.50 % |
| Linear Baseline | 86.01 % |
| Logistic 1024 | 92.30 % |
| Logistic 256 | 93.55 % |
| Ensemble | 94.05 % |

Table 4: F-score

| Model | TP | FP | FN | Precision | Recall | F-score |
|---|---|---|---|---|---|---|
| DCASE CNN | 33 | 62 | 197 | 34.73 % | 14.34 % | 20.31 % |
| DCASE Template Matching | - | - | - | 1.08 % | 14.46 % | 2.01 % |
| Logistic Baseline | 21 | 1333 | 209 | 1.55 % | 9.13 % | 2.65 % |
| Ensemble | 11 | 1871 | 219 | 0.58 % | 4.78 % | 1.042 % |

## 4.2 Baseline Model

### 4.2.1 DCASE model implementation

The original model implements prototypical networks using a deep learning approach designed for few-shot scenarios. It incorporates a non-linear mapping from the input space to embedding space, which is learnt using a convolutional neural network (CNN). Then, the class prototype is calculated by taking a mean of its support set in the embedding space. Finally, classification of a query point is conducted by finding the nearest class prototype.

### 4.2.2 Our model

Instead of using a CNN to learn the embedding, we explored the following two models: slightly modified logistic regression (1) and linear regression (2). During training part of training set is set out for validation. The validation accuracy is reported on this set and is not to be confused with the validation set provided by DCASE. The validation set of DCASE is used to calculate precesion, recall and F-score. We found that the logistic regression had a higher validation accuracy than the liner regression because it could model non-linear relationships. We found that we could improve the validation accuracy more by increasing the learning rate and creating an ensemble model consisting two logistic layers. The input dimension is $17 \times 128$, where 17 is the size of equal length segments and 128 is the number of MEL features. The input is flattened and fed to the model. The output dimensions of the two logistic layers are either 1024 or 256, respectively (based on hyperparameter search). The models were trained for 15 epochs with a learning rate of 0.0001. The validation accuracy of each of our models is shown in Table 3.

$$F_A(x) = \frac{1}{1 + \exp(Ax)} \qquad (1)$$

$$F_A(x) = Ax \qquad (2)$$

### 4.2.3 NumPy Implementation

The implementation provided by DCASE utilized PyTorch for training a CNN, but we developed a working implementation of our models (linear and logistic regression) in NumPy. The major part of the conversion from PyTorch to NumPy consisted of the backpropagation algorithm with the Prototypical loss function [8].

Given a training episode (batch), the episode is divided into support $S_k$ and query $Q_k$ set ($k$ denotes a particular class) to mimic the few shot scenario. We used 10-way 5-shot learning task during training procedure. Class prototypes are calculated as the mean of the embeddings from the support set points:

$$\mathbf{c}_k = \frac{1}{|S_k|} \sum_{(\mathbf{x}_i, y_i) \in S_k} F_A(\mathbf{x}_i) \qquad (3)$$

The negative distance of the embeddings of the query points from the prototypes can be used to assign probabilities to the classes for query set (large negative distance (close points) $\implies$ high probability). For $\mathbf{x} \in Q_k$, softmax is used to find the probability of belonging to a particular class

$$p_A(y = k|\mathbf{x}) = \frac{\exp(-d(F_A(\mathbf{x}), \mathbf{c}_k))}{\sum_{k'} \exp(-d(F_A(\mathbf{x}), \mathbf{c}_{k'}))} \qquad (4)$$

The loss function is given by:
$$J(A) = -\log\ p_A(y = k|\mathbf{x}) \qquad (5)$$

In our implementation, instead of using the negative log likelihood over the softmax function, we use softmax cross-entropy loss which is theoretically equivalent but easier to implement. The following influence diagram was used for backpropagation algorithm (**x** represents an episodic batch) -

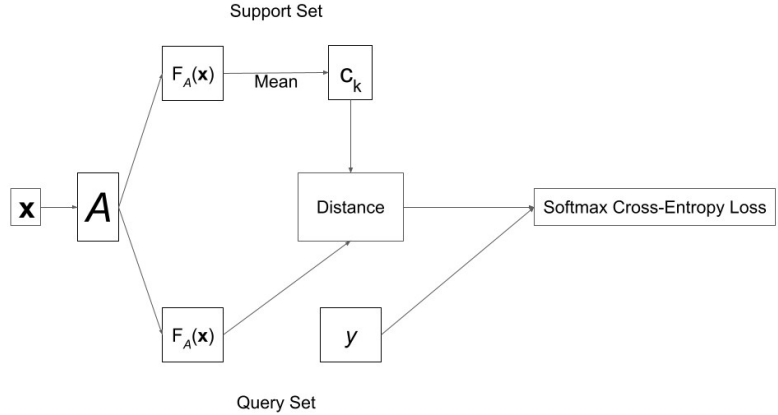

### 4.3 Ensemble Methods

Ensemble methods is a machine learning technique that combines several base models in order to produce one optimal predictive model.

In our implementation, we combined both the logistic regression methods to perform ensemble learning (i.e. the logistic regression with output dimension of 256 and logistic regression with output dimension of 1024).

It can be seen in Table 3 that the ensemble model achieved a validation accuracy of $94.05$ %, thereby performing better than both of the individual the logistic models.

Additionally, we tried a linear + logistic ensemble model, but the validation accuracy was quite poor. This is because our data is non-linear, and the linear regression performs poorly on this type of data.

Although, the ensemble model performs better in terms of Validation accuracy, it does not achieve an F-score similar to that achieved by the DCASE CNN. Thus, one can conclude that the accuracy of the model is not directly related with F-score. It can be seen in Table 4 that our final ensemble model had a large number of false positives (FP). This indicates that the model over-predicted the number of events in the validation set audio. This model also achieved a precision which was much lower than that achieved by the DCASE CNN. This further indicates that our model is too sensitive to ambient sounds, like wind, in the field recordings. Furthermore, the model is unable to differentiate between these noises and the animal vocalizations.

### 4.4 Kernel Methods

For introducing non-linearity into the model, we explored various kernel functions which can be used as part of prototypical loss function. This is achieved by rewriting the formula for euclidean distance as follows:

$$||x - y||^2 = ||x||^2 + ||y||^2 - 2xy = K(x,x) + K(y,y) - 2K(x,y) \quad (6)$$

We specifically tried two kernel functions:

#### 4.4.1 Dot product with positive definite matrix using Cholesky decomposition

Kernel function is given by

$$K(\mathbf{x}, \mathbf{y}) = \mathbf{x}^T Q \mathbf{y} \quad (7)$$

where the matrix Q is a positive definite matrix. Cholesky decomposition splits any positive definite matrix Q as

$$Q = LL^T \quad (8)$$

where $L$ is a lower triangular matrix. This can further be rewritten as-

$$Q = LDL^T \quad (9)$$

where $L$ is a lower triangular matrix with all diagonal elements as 1 and D is positive definite diagonal matrix. In our case we set all the diagonal elements of $D$ to be equal to one. During training, we also update the matrix $L$ and mask out the upper triangular elements after each update. The model is then trained for 15 epochs with a learning rate of 0.0001. The best validation accuracy achieved was 93.23%. Table 6 compares the precision, recall and f-score with CNN implementation.

#### 4.4.2 RBF kernels

The kernel function for RBF is given by

$$k(\mathbf{x}, \mathbf{y}) = e^{-\gamma ||\mathbf{x} - \mathbf{y}||^2} \quad (10)$$

where $\gamma$ is a hyperparameter which has to be tuned. We try out four values of $\gamma$ specifically $\gamma \in 0.001, 0.01, 0.1, 1$. Each model is trained for 15 epochs with a learning rate of 0.0001. Table 6 compares the validation accuracy, precision, recall and f-score with CNN implementation.

Table 5: Comparison between kernel implementation and DCASE CNN model

| Model | Accuracy | Precision | Recall | F-score |
|---|---|---|---|---|
| DCASE CNN | 98.55% | 34.73% | 14.34% | 20.308 |
| Cholesky decomposition | 93.23% | 0.4% | 3.91% | 0.719 |
| RBF ($\gamma = 0.5$) | 77.17% | 0.31% | 23.91% | 0.606 |
| RBF ($\gamma = 0.01$) | 89.78% | 0.52% | 36.96% | 1.028 |
| RBF ($\gamma = 0.1$) | 83.25% | 0.33% | 40% | 0.662 |
| RBF ($\gamma = 0.001$) | 79.98% | 0.43% | 40% | 0.843 |

Table 6: Comparison between modified loss function model and DCASE CNN model

| Model | Accuracy | Precision | Recall | F-score |
|---|---|---|---|---|
| DCASE CNN | 98.55% | 34.73% | 14.34% | 20.308 |
| Prototypical+distance loss | 90.94% | 1.56 % | 8.26% | 2.619 |

### 4.5 Maximizing the distance between prototypes

Prototypical loss tries to minimize the euclidean distance between the internal representations of a sample event and its corresponding prototype. We can further extend this loss to train the model to learn an internal representation such that the euclidean distance between the prototypes of different classes are maximized. This is achieved by penalising the model when the distance between any two prototypes is lesser than a distance $\delta_v$ [3]. The modified loss is given by:

$$\text{Loss} = \text{Prototypical loss} + \lambda \sum_{i \neq j} max(0, \delta_v - ||\mu_i - \mu_j||^2) \quad (11)$$

where $\lambda$ is a hyperparameter which can be tuned to adjust the relative importance of prototypical loss and the distance loss. The logistic model is then trained for 30 epochs with a learning rate of 0.0001, $\lambda = 0.1$ and $\delta_v = 10$. Table 6 compares the results of the trained model with DCASE CNN model.

## 5 Discussion and Analysis

While it is possible to implement few-shot bioacoustic event detection using only machine learning methods, this approach has some trade-offs in comparison to a deep learning solution. In general, it was found that the logistic regression models greatly over-predicted the number of events given the validation set. This lack of precision can in-part be attributed to the need to flatten the input data, resulting in the loss of temporal information. The CNN solutions were able to preserve the temporal information in the data and, therefore, achieved a higher precision.

As few-shot learning is most commonly used in biological studies, imbalanced class distribution will undoubtedly exist. Therefore, the F-score is a more useful metric for evaluation. In other words, the F-score is used when it is crucial to minimize the number of false negatives and false positives in the predictions [5]. Alternatively, accuracy is a useful evaluation metric when it is most important to capture all of the true positive and true negative events [5]. Thus, while our logistic models achieved a high validation accuracy during training, this did not correlate to a sufficient F-score. Overall, while our model was able to correctly predict the majority of animal vocalizations in the recordings, it also predicted many events that are not truly of interest.

While our model was unable to achieve the same F-score as the DCASE CNN solution, we were able to determine that a single logistic regression outperforms template matching and linear regression, as demonstrated by the achieved F-score in Table 4. Our results also indicate why, in some cases, deep learning is better suited for classification. Finally, when trying to improve this model, we found that the Cholesky decomposition and RBF kernel gave similar results. In the future, it would be interested to explore embedding propagation [6], which smooths the manifold, to see if this could yield improved results.